\documentclass[a4paper,english,twocolumn]{svjour3}
\usepackage[T1]{fontenc}
\usepackage[latin9]{inputenc}
\usepackage{babel}
\usepackage{array}
\usepackage{textcomp}
\usepackage{url}
\usepackage{multirow}
\usepackage{amsmath}
\usepackage{graphicx}
\usepackage[authoryear]{natbib}
\usepackage{subscript}
\usepackage[unicode=true,pdfusetitle,
 bookmarks=true,bookmarksnumbered=false,bookmarksopen=false,
 breaklinks=true,pdfborder={0 0 0},pdfborderstyle={},backref=false,colorlinks=false]
 {hyperref}
\usepackage{breakurl}

\makeatletter

\providecommand{\tabularnewline}{\\}

\usepackage{marvosym}
\newcommand{\envelope}{(\raisebox{-.5pt}{\scalebox{1.45}{\Letter}}\kern-1.7pt)}

\authorrunning{Huang et al.}

\makeatother

\begin{document}

\title{Propagation and synchronization of reverberatory bursts in developing
cultured networks}

\author{Chih-Hsu Huang$^{1}$ \and Yu-Ting Huang$^{1,2}$ \and Chun-Chung
Chen$^{1}$ \and\\
 C. K. Chan$^{1,2}$}

\institute{\Letter  ~Chun-Chung Chen \at cjj@phys.sinica.edu.tw \and\\
\\
$^{1}$ Institute of Physics, Academia Sinica\\
Nangang, Taipei, Taiwan 115, ROC\\
\\
$^{2}$ Department of Physics and Center for Complex Systems\\
National Central University\\
Chungli, Taiwan 320, ROC}
\maketitle
\begin{abstract}
Developing networks of neural systems can exhibit spontaneous, synchronous
activities called neural bursts, which can be important in the organization
of functional neural circuits. Before the network matures, the activity
level of a burst can reverberate in repeated rise-and-falls in periods
of hundreds of milliseconds following an initial wave-like propagation
of spiking activity, while the burst itself lasts for seconds. To
investigate the spatiotemporal structure of the reverberatory bursts,
we culture dissociated, rat cortical neurons on a high-density multi-electrode
array to record the dynamics of neural activity over the growth and
maturation of the network. We find the synchrony of the spiking significantly
reduced following the initial wave and the activities become broadly
distributed spatially. The synchrony recovers as the system reverberates
until the end of the burst. Using a propagation model we infer the
spreading speed of the spiking activity, which increases as the culture
ages. We perform computer simulations of the system using a physiological
model of spiking networks in two spatial dimensions and find the parameters
that reproduce the observed resynchronization of spiking in the bursts.
An analysis of the simulated dynamics suggests that the depletion
of synaptic resources causes the resynchronization. The spatial propagation
dynamics of the simulations match well with observations over the
course of a burst and point to an interplay of the synaptic efficacy
and the noisy neural self-activation in producing the morphology of
the bursts.

\end{abstract}

\section{Introduction}

During the development of neural systems, spontaneous and synchronous
activities can appear following the outgrowth of neurites and before
the availability of external stimulus inputs \citep{segev_formation_2003,meister_synchronous_1991}.
These activities are believed to play an important role in the formation
and organization of functional neural circuitries \citep{katz_synaptic_1996,turrigiano_homeostatic_2004,harris_neural_1981,crair_neuronal_1999}.
The investigation of these network activities can help to elucidate
the cellular and network mechanisms involved in neural development
\citep{zhang_electrical_2001,bi_synaptic_2001,blankenship_mechanisms_2010,kerschensteiner_spontaneous_2014}
and will lead to a better understanding of the functioning of a brain
\citep{penn_brain_1999,hua_neural_2004,chiappalone_dissociated_2006,pu_developing_2013}.
Among approaches to study the spontaneous activity of developing neural
systems, dissociated cultures of cortical or hippocampal neurons on
a multi-electrode array (MEA) have been used for decades as experimental
models for observing the dynamics of growing networks \citep{thomas_miniature_1972,pine_recording_1980,gross_recording_1982,potter_new_2001}.

Usually, spontaneous activities can be observed after about a week
in vitro and the activities are later synchronized into episodic network
bursts \citep{maeda_mechanisms_1995,chiappalone_dissociated_2006}.
Interesting patterns of these neural bursts have been reported \citep{van_pelt_longterm_2004,wagenaar_extremely_2006,raichman_identifying_2008},
where the activity level of firing rate in the burst can have repeated
peaks of rise-and-falls called reverberations at a time scale of hundreds
of milliseconds following the initial spike of activities \citep{lau_synaptic_2005}.
These so-called ``super bursts'' \citep{wagenaar_extremely_2006}
can last for seconds and, for their similarity in the time scales,
are thought to be important to understand cognitive functions such
as working memory on the cellular and network levels \citep{wang_synaptic_2001,lau_synaptic_2005,compte_computational_2006,mongillo_synaptic_2008,volman_synaptic_2011,bermudezcontreras_formation_2013,dranias_short-term_2013}.

There have been active studies on the initiation of in vitro neural
bursts \citep{feinerman_identification_2007,eckmann_leader_2008}
focusing on both the role of hub neurons \citep{cossart_operational_2014,schroeter_emergence_2015}
and topological effects \citep{orlandi_noise_2013}. The development
of high-density MEA systems has enabled more detailed investigation
of the activity propagation in the neural bursts. Notably, the collective
dynamics of spiking neurons such as center-of-activity trajectory
(CAT) allow the identification of a propagation phase and a reverberation
phase in the progression of a burst event \citep{gandolfo_tracking_2010}.

In the current study, we use a similar high-density MEA system to
investigate reverberatory bursts observed in the development of dissociated
cortical cultures. Instead of considering reduced dynamics such as
principal components or CAT, we use a propagation model to predict
the location of each occurring spike. The effectiveness of such prediction
allows the classification of the spikes into \emph{evoked} and \emph{spontaneous}
ones, and can be used in reverse for an inference on the spreading
speed of the recorded spiking activity. We find a recovering dominance
of the evoked spikes over the reverberatory phase of a burst following
their reduction after the initial propagating wave.

We implement a physiologically realistic model of neuronal systems
\citep{volman_calcium_2007} on a geometrically-constrained, two-dimensional
network and identify sets of parameters that can produce reverberatory
bursts qualitatively similar to the experimental observations. With
all dynamical variables being available in computer simulations, we
clarify the roles played by the neuronal noise as well as the depletion
of synaptic resources in the continuation and termination of the reverberatory
bursts. We find that the depletion, which is responsible for terminating
the burst events \citep{cohen_network_2011}, is also important in
restoring the synchrony of reverberatory activity during the bursts.

\section{Materials and methods}

\subsection{Cell cultures and experimental setup}

Cortical neurons were dissociated from Wistar rat at embryonic day
17 (E17). Tissues were digested by 0.125\% trypsin and plated on the
BioChip 4096E (3Brain, Switzerland) previously coated with poly-D-lysine
(0.1mg/ml) and laminin (0.1mg/ml) to promote the adhesion of neurons.
About 6\texttimes 10$^{4}$ neurons were plated, completely covering
an active area of 6\texttimes 6 mm\textsuperscript{2}, yielding a
density of the culture of about 1.7\texttimes 10$^{3}$ neurons/mm\textsuperscript{2}.
Cultures were filled with 1 mL culture medium at 30 min after plating
and incubated at 37\textcelsius{} in the presence of 5\% CO\textsubscript{2}.
Half of the medium was refreshed twice a week.

\subsection{Electrophysiological signals}

Electrophysiological activities of neurons were recorded with the
original culture medium once every other day since 6 DIV in 5\% CO\textsubscript{2}
at room temperature (24\textcelsius ). Before recording, the culture
was kept at room temperature for 10 min for stabilization and placed
back to the incubator immediately after the recording for future measurement.
The chip 4096E has a recording area of 5.12\texttimes 5.12 mm\textsuperscript{2}
covered by 64\texttimes 64 electrodes. The area of each electrode
is 21\texttimes 21 $\mu$m\textsuperscript{2} with an inter-electrode
separation of 81$\mu$m.

\begin{figure}
\includegraphics[width=1\columnwidth]{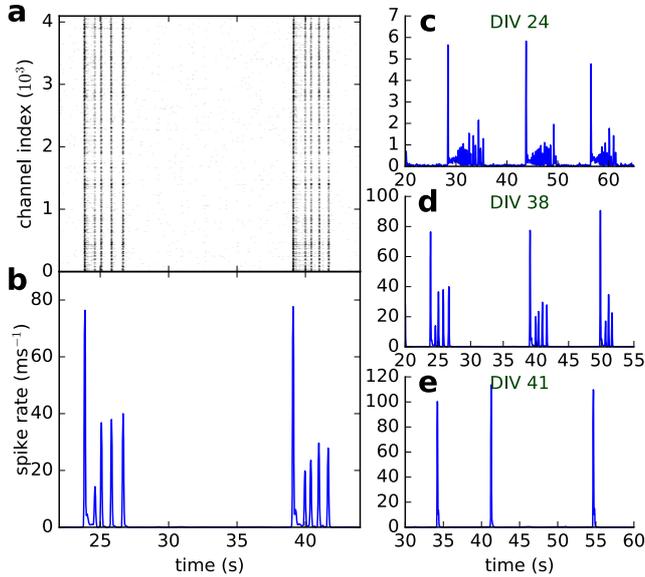}

\caption{\textbf{a} Raster plot of detected spikes from the culture E at DIV
38 and \textbf{b} the corresponding time histogram of firing rate.
\textbf{c} to\textbf{ e} Firing-rate histograms for different DIVs
(as labeled) of the culture E\label{fig:Raster-Histograms}}
\end{figure}

The network activity was acquired at a sampling rate of 7.7 kHz for
each electrode. Each recording data set includes network activity
of 5 min. But, the data sets containing unstable activity patterns,
long silent periods, or abnormal activities with, e.g., strong noise,
were excluded for further processing. The qualified data sets for
further processing are listed in Table \ref{tab:selected-cultures}.
\begin{table}[ht]
\caption{List of experimental recordings\label{tab:selected-cultures}}

\centering{}%
\begin{tabular}{ccc>{\centering}m{2.2cm}}
\textbf{Culture}  & \textbf{DIV}  & \textbf{Reverberation}  & \textbf{Spreading Speed (mm/s)}\tabularnewline
\hline 
\multirow{2}{*}{A} & 12  & No  & 93.15\tabularnewline
 & 25  & Yes  & 140.9\tabularnewline
\hline 
\multirow{2}{*}{B} & 13  & No  & 30.78\tabularnewline
 & 25  & Yes  & 75.33\tabularnewline
\hline 
\multirow{3}{*}{C} & 33  & Yes  & 200.9\tabularnewline
 & 40  & Yes  & 179.8\tabularnewline
 & 68  & No  & 554.9\tabularnewline
\hline 
\multirow{4}{*}{D} & 26  & No  & 145.0\tabularnewline
 & 33  & Yes  & 458.5\tabularnewline
 & 40  & Yes  & 480.3\tabularnewline
 & 68  & No  & 517.6\tabularnewline
\hline 
\multirow{3}{*}{E} & 24  & Yes  & 77.76\tabularnewline
 & 38  & Yes  & 435.0\tabularnewline
 & 41  & No  & 398.5\tabularnewline
\hline 
\multirow{3}{*}{F} & 12  & Yes  & 116.6\tabularnewline
 & 39  & Yes  & 403.4\tabularnewline
 & 54  & No  & 448.7\tabularnewline
\hline 
\end{tabular}
\end{table}
 Spontaneous activities can be observed after about 2 weeks in vitro,
comprising isolated spikes and short bursts involving many neurons
(electrodes), e.g., the one shown in Fig.~\ref{fig:Raster-Histograms}.
 The isolated spikes produced in neurons are detected by the BrainWave
software through the Precise Timing Spike Detection with threshold
values that are 8 times of the standard deviation of spike-free signals.

\subsection{Detection of bursts and activity peaks}

The bursts are detected as follows. 
\begin{figure}
\includegraphics[width=1\columnwidth]{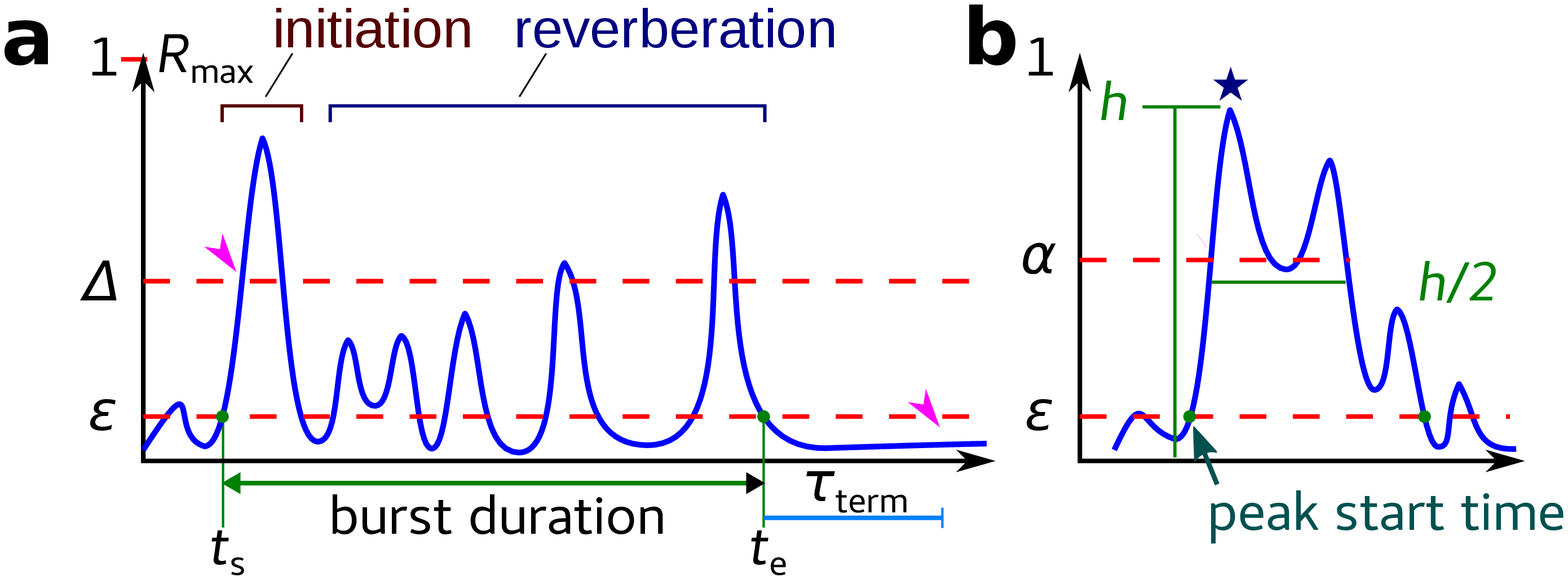}

\caption{Illustrated parameters for the detection of \textbf{a} a burst and\textbf{
b} a spike rate peak (marked by the star) as described in the text.
The vertical axes are spike rate in units of their maximum $R_{\text{max}}$
of the recording while the horizontal are time axes \label{fig:Illustrated-parameters}}
\end{figure}
 The spike rate $R\left(t\right)$ at each instance $t$ is measured
as an average over the time window of size $\lambda$ centered at
the time. For detected spike times from each MEA recording, typically
of a 5-minute duration, the maximum of the spike rate $R_{\text{max}}$
is first determined. A lower threshold $R_{\text{lower}}\equiv\epsilon R_{\text{max}}$
is used to decide whether the culture is in an active state as illustrated
in Fig. \ref{fig:Illustrated-parameters}a. A reverberatory burst
typically starts with a strong activity peak for the initiation phase
followed by varying activity level or peaks in the reverberation phase
as illustrated in Fig. \ref{fig:Illustrated-parameters}a. A burst
is registered starting at $t_{s}$ when the culture becomes active
at $t_{s}$ and stays active until the spike rate reaches an upper
threshold of $R_{\text{upper}}\equiv\Delta R_{\text{max}}$. The registered
burst ends at $t_{e}$ when the culture becomes inactive at $t_{e}$
and stays inactive at least for a duration of $\tau_{\text{term}}$.
The empirical values for the parameters used in the burst detection
of both the experimental and simulated data are: $\lambda=0.02$s,
$\epsilon=0.04$, $\Delta=0.2$, and $\tau_{\text{term}}=1.5$s.

The peaks or reverberations could also be identified using the same
method as described above with a different set of empirical parameter
values. However, here we use a simpler definition that is time-symmetric:
A peak is defined as a significant maximum (height $h$ > $\alpha R_{\text{max}}$
> $R_{\text{lower}}$) in the firing rate of a continuous time interval
where the rate is above half of this maximum firing rate as marked
in Fig. \ref{fig:Illustrated-parameters}b. Preceding this interval
and following the previous peak, if the firing rate of the culture
stays above the lower threshold, the minimum of firing rate is considered
the starting time of this peak. Otherwise, the starting time is registered
as the time when the rate crosses the lower threshold. The state variables
of the system representing the internal noise and degree of depletion,
which are only available in simulation results, are determined at
the start time of a peak to correlate with the characteristics of
the peak.

\subsection{Activity propagation and predictability of spiking electrodes}

\begin{figure}
\includegraphics[width=1\columnwidth]{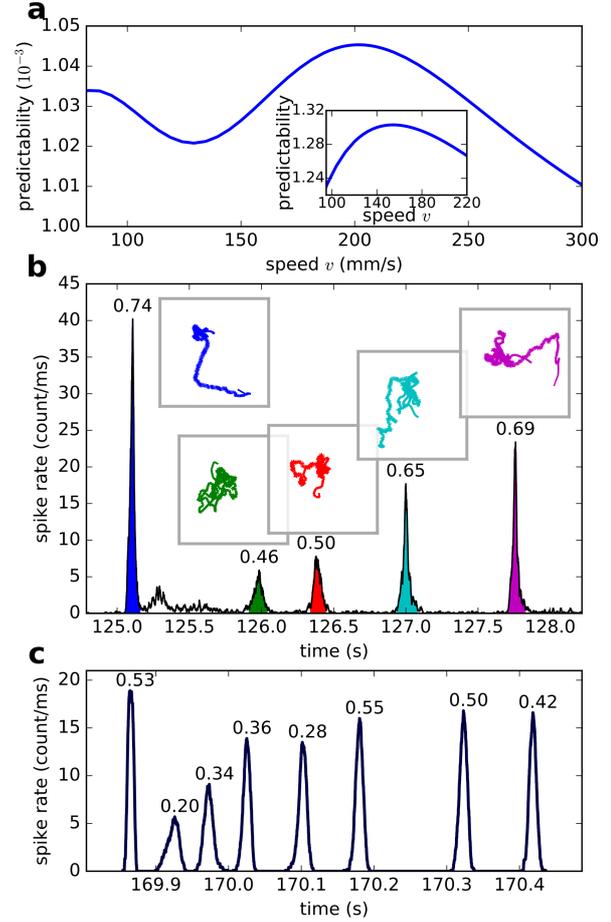}

\caption{\textbf{a} Predictability as a function of presumed spreading speed
for culture C at 33 DIV. Inset is a similar plot from a simulation.
\textbf{b} Time-histogram of a typical reverberatory burst with identified
activity peaks color coded with corresponding center-of-activity trajectories
in the insets. The numbers next to the rate peaks show the fractions
of the evoked spikes out of all spikes in the activity peaks. \textbf{c}
Time-histogram for a simulated reverberatory burst. The fractions
of evoked spikes are similarly labeled for each activity peak \label{fig:Predictability-Histogram}}
\end{figure}

The propagation of the spiking activity in a burst can be visually
observed from the animated replay of sustaining spikes (Online Resource).
To quantify the wave-like propagation of the initial sweep of activity
and the subsequent distributed activation of neurons, we introduce
a simple linear-spread diffusive model that can be used to predict
the electrode for the next spike using spikes that have already been
recorded. The probability for the next spike occurring at time $t$
to be on the electrode at $\mathbf{r}$ is given by 
\begin{equation}
P\left(\mathbf{r}\right)=\frac{1}{N}\sum_{\substack{\{i|t_{i}<t\}}
}e^{-\left(t-t_{i}\right)/\tau_{p}}\frac{1}{L_{i}^{2}}e^{-\left|\mathbf{r}-\mathbf{r}_{i}\right|^{2}/L_{i}^{2}}\label{eq:linear-spread-diffusive}
\end{equation}
where $t_{i}$ and $\mathrm{r}_{i}$ are time and location of the
previous spike $i$, $L_{i}\equiv v\left(t-t_{i}\right)$ is the spreading
influence range of the spike $i$, $\tau_{p}$ is the decay time of
the influence, $N\equiv\sum_{\mathbf{r}}P\left(\mathbf{r}\right)$
is the normalization factor, and $v$ is the spreading speed of the
influence. We note that the probability (\ref{eq:linear-spread-diffusive})
is conditional on a spike occurring at time $t$, and should be multiplied
with the spike rate $R\left(t\right)$ for predicting the occurrence
of a spike at $\mathbf{r}$. We define the predictability of spikes
as the average of $P\left(\mathbf{r}\right)$ over all spikes in a
recording comparing to the uniform distribution, which tells us how
well the location of a spike can be predicted from previous spikes
using the simple model (\ref{eq:linear-spread-diffusive}). For each
recording, we find the value of $v$ that maximizes the predictability
relative to a surrogate with randomized spike positions as shown in
Fig.~\ref{fig:Predictability-Histogram}a and these values are included
in Table (\ref{tab:selected-cultures}) for all recordings. 
 With the optimal value $v$, a spike is considered an evoked spike
if its position $\mathbf{r}$ satisfies $P\left(\mathbf{r}\right)>2P_{0}$
where $P_{0}=1/N_{\text{elec}}\approx2.4\times10^{-4}$ is the average
probability for the spike to occur at an electrode out of the $N_{\text{elec}}=4096$
electrodes for our MEA. The number ratio of evoked spikes to the total
spikes within the rate peaks of a burst are shown next to the corresponding
peaks in Fig.~\ref{fig:Predictability-Histogram}b.

\subsection{Computer simulations}

To gain insight into the dynamics of the reverberatory bursts, we
use a neuronal synaptic model similar to that described by Volman
et al \citep{volman_calcium_2007}. The model uses Morris\textendash Lecar
(ML) \citep{morris_voltage_1981} neurons connected with Tsodyks\textendash Markram
(TM) \citep{tsodyks_neural_1997} synapses. The dynamics of neurons
are governed by the ML equations,\begin{subequations}\label{subsec:Morris-Lecar}
\begin{eqnarray}
C\frac{dV}{dt} & = & -I_{{\rm ion}}+G\left(V_{r}-V\right)+I_{{\rm bg}},\\
\frac{dW}{dt} & = & \theta\frac{W_{\infty}-W}{\tau_{W}}
\end{eqnarray}
\end{subequations}where 
\begin{equation}
I_{{\rm ion}}=g_{{\rm Ca}}m_{\infty}\left(V-V_{{\rm Ca}}\right)+g_{{\rm K}}W\left(V-V_{{\rm K}}\right)+g_{{\rm L}}\left(V-V_{{\rm L}}\right)
\end{equation}
is the current through the membrane ion channels,\begin{subequations}
\begin{eqnarray}
\tau_{W} & = & \left(\cosh\frac{V-V_{3}}{2V_{4}}\right)^{-1},\\
W_{\infty} & = & \frac{1}{2}\left(1+\tanh\frac{V-V_{3}}{V_{4}}\right),\\
m_{\infty} & = & \frac{1}{2}\left(1+\tanh\frac{V-V_{1}}{V_{2}}\right)
\end{eqnarray}
\end{subequations}are the voltage dependent dynamic parameters, and
the threshold $V_{{\rm th}}$ of membrane potential defines the spiking
events which result in synchronous releases of neural transmitters
at the efferent synapses. Additionally, a residual calcium variable
$R_{{\rm Ca}}$ driven by the spiking events,
\begin{equation}
\frac{d}{dt}R_{{\rm Ca}}=\frac{-\beta R_{{\rm Ca}}^{n}}{k_{R}^{n}+R_{{\rm Ca}}^{n}}+I_{p}+S\gamma\log\frac{R_{{\rm Ca}}^{0}}{R_{{\rm Ca}}},
\end{equation}
where the spike train is $S=\sum_{\sigma}\delta\left(t-t_{\sigma}\right)$
with $t_{\sigma}$ being the time of the spike event $\sigma$, is
used to determine the rate,
\begin{equation}
\eta=\eta_{{\rm max}}\frac{R_{{\rm Ca}}^{m}}{k_{a}^{m}+R_{{\rm Ca}}^{m}},\label{eq:asynchro-rate}
\end{equation}
of synapse-dependent asynchronous releases of neural transmitters
(see below) following an independent Poisson process at each efferent
synapse. The neural transmitters released by the spike-driven synchronous
and calcium-dependent asynchronous events follow a four-state decaying
dynamics based on a modification of the TM model,\begin{subequations}\label{subsec:TUMs-dynamics}
\begin{eqnarray}
\frac{dX}{dt} & = & \frac{Q}{\tau_{s}}+\frac{Z}{\tau_{r}}-uXS-X\xi\\
\frac{dY}{dt} & = & -\frac{Y}{\tau_{d}}+uXS+X\xi\label{eq:active-state}\\
\frac{dZ}{dt} & = & \frac{Y}{\tau_{d}}-\frac{Z}{\tau_{r}}-\frac{Z}{\tau_{l}}\\
\frac{dQ}{dt} & = & \frac{Z}{\tau_{l}}-\frac{Q}{\tau_{s}}.\label{eq:super-inactive-state}
\end{eqnarray}
\end{subequations}where $\xi=\bar{\xi}\sum_{a}\delta\left(t-t_{a}\right)$
summing over the asynchronous release events $a$ with a Poisson rate
given by (\ref{eq:asynchro-rate}), to include a super-inactive state
$Q$. Multiplying by the synaptic weights, the fractions of neural
transmitters in the active state $Y$ (\ref{eq:active-state}) determine
the contribution of the afferent synapses to the membrane conductance
$G$ of a post-synaptic neuron through a linear sum 
\begin{equation}
G_{i}=\sum_{j}w_{ji}Y_{ji}
\end{equation}
over all pre-synaptic neurons $j$ of the given post-synaptic neuron
$i$. Following \citet{volman_calcium_2007}, the synaptic weights
$w$ are randomly drawn from a truncated Gaussian distribution with
a width that is $\pm20\%$ of its mean $\bar{w}$ for the connected
neurons.

\begin{figure}
\includegraphics[width=1\columnwidth]{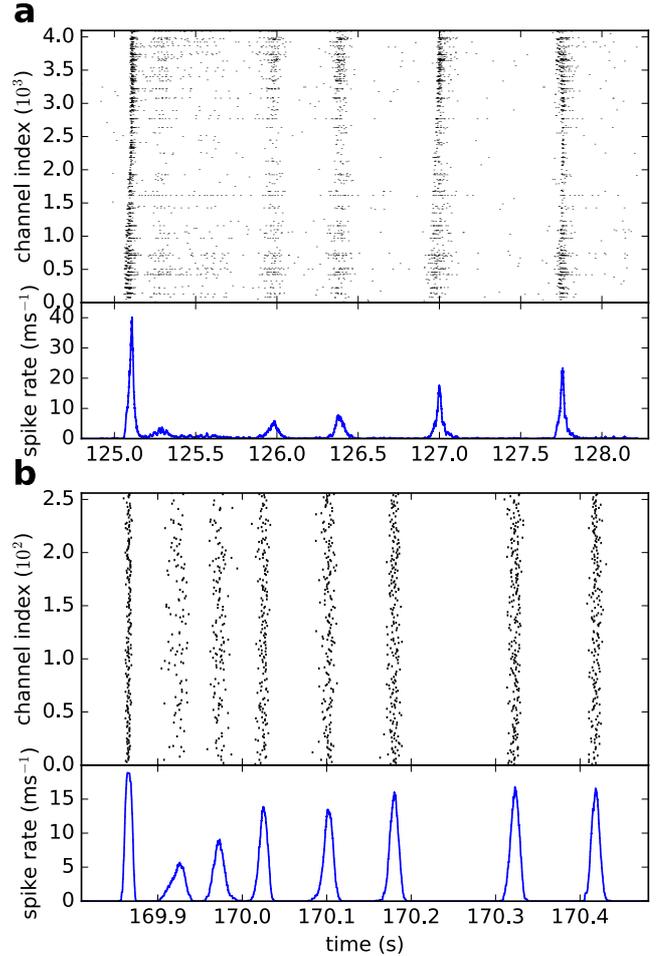}

\caption{\textbf{a} Raster plot of recorded spikes and corresponding spike-rate
histogram for the bursting event shown in Fig. \ref{fig:Predictability-Histogram}b.
\textbf{b} Raster plot of simulated spikes and corresponding histogram
for the bursting event shown in Fig. \ref{fig:Predictability-Histogram}c\label{fig:raster plots}}
\end{figure}

We place the model neurons on a 2D geometrical network with connection
probability between two neurons decaying exponentially with the distance
between them. Most of the model parameters used in our simulations
follow the values given in \citep{volman_calcium_2007} and can be
found in Table~\ref{tab:model-parameters}.
\begin{table}
\caption{Values of parameters used in simulations.\label{tab:model-parameters}}

\begin{centering}
\begin{tabular*}{0.85\columnwidth}{@{\extracolsep{\fill}}rlrlrl}
\hline 
\multicolumn{6}{c}{Morris\textendash Lecar model}\tabularnewline
\hline 
$V_{{\rm Ca}}$ & $100$ ${\rm mV}$ & $V_{2}$ & $15$ ${\rm mV}$ & $g_{{\rm L}}$ & $0.5$ ${\rm mS}$\tabularnewline
$V_{{\rm K}}$ & $-70$ ${\rm mV}$ & $V_{3}$ & $0$ ${\rm mV}$ & $C$ & $1$ ${\rm \mu F}$\tabularnewline
$V_{{\rm L}}$ & $-65$ ${\rm mV}$ & $V_{4}$ & $30$ ${\rm mV}$ & $\text{\ensuremath{\theta}}$ & $0.2$ ${\rm ms}^{-1}$\tabularnewline
$V_{r}$ & $0$ ${\rm mV}$ & $g_{{\rm Ca}}$ & $1.1$ ${\rm mS}$ & $V_{{\rm th}}$ & $10$ ${\rm mV}$\tabularnewline
$V_{1}$ & $-1$ ${\rm mV}$ & $g_{{\rm K}}$ & $2$ ${\rm mS}$ &  & \tabularnewline
\hline 
\end{tabular*}
\par\end{centering}
\begin{centering}
\begin{tabular*}{0.8\columnwidth}{@{\extracolsep{\fill}}rlrlrl}
\hline 
\multicolumn{6}{c}{Tsodyks\textendash Markram synaptic transmission}\tabularnewline
\hline 
$\tau_{d}$ & $10$ ${\rm ms}$ & $\tau_{l}$ & $800$ ${\rm ms}$ & $u$ & $0.25$\tabularnewline
$\tau_{r}$ & $250$ ${\rm ms}$ & $\tau_{s}$ & $5000$ ${\rm ms}$ & $\bar{\xi}$ & $0.02$\tabularnewline
\hline 
\end{tabular*}
\par\end{centering}
\centering{}%
\begin{tabular*}{0.95\columnwidth}{@{\extracolsep{\fill}}rlrlrl}
\hline 
\multicolumn{6}{c}{Residual calcium dynamics}\tabularnewline
\hline 
$\beta$ & $0.005$ $\frac{{\rm \mu M}}{{\rm ms}}$ & $\gamma$ & $0.033$ & $k_{a}$ & $0.13$ ${\rm \mu M}$\tabularnewline
$k_{R}$ & $0.4$ ${\rm \mu M}$ & $R_{{\rm Ca}}^{0}$ & $2000$ ${\rm \mu M}$ & $m$ & $4$\tabularnewline
$I_{p}$ & $1.1\times10^{-4}$ $\frac{{\rm \mu M}}{{\rm ms}}$ & $\eta_{{\rm max}}$ & $0.32$ ${\rm ms}^{-1}$ & $n$ & $2$\tabularnewline
\hline 
\end{tabular*}
\end{table}
 The time constants of TM dynamics, background currents for ML neurons,
and synaptic weights are adjusted uniformly to reach simulated time-histograms
that qualitatively reproduce the experimental results as seen in Fig.
\ref{fig:Predictability-Histogram}. The raster plots for the simulated
burst and the experimentally observed burst in Fig. \ref{fig:Predictability-Histogram}
are shown in Fig. \ref{fig:raster plots}. 
 For current study, we focus on the reverberatory bursts with distinct
reverberation peaks or sub-bursts in the spike rate histogram.

The same burst and peak detections for the experimental measurements
are applied to the simulation results with slightly different empirical
parameters. Comparing to the experiments, the full dynamics of the
simulations is readily available as numerical data and can be further
analyzed to clarify the physical mechanisms of the bursting behavior.
Beside recording the time and neuron of each spike for the calculation
of a time-histogram and keeping track of activity propagation, we
are interested in the information of neuronal noise and the depletion
of synaptic resources. The former is represented by the average concentration
of residual calcium that governs the asynchronous release while the
later is represented by the average fraction of inactive and super
inactive neural transmitters which deplete the available neural transmitters
in a bursting cycle. Both of the values are retained at the start
time of each detected peak in the spike-rate histogram and used to
correlate with the properties of each peak.

\subsection{Implementations}

We implemented the computational model in the C++ programming language
using the Common Simulation Tools framework. The simulation codes
along with the framework are included in the supplementary materials
of the paper. The spike data from the MEA recordings as well as the
computer simulations were processed with the Python3 programming language
and most of the data plots were produced using the Matplotlib library
module. A Jupyter Notebook containing the Python3 codes for data processing
and plotting is also included in the supplementary materials.

\section{Results}

\begin{figure}
\includegraphics[width=8cm]{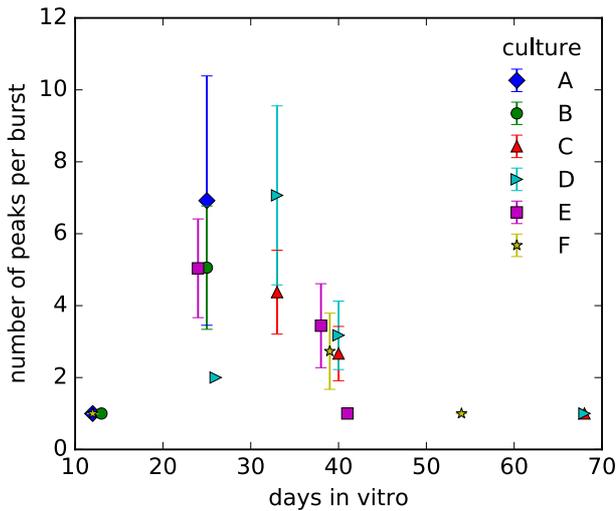}

\caption{Number of detected peaks in spike rate per burst as a function of
DIV of the cultures. The reverberation of the bursts is maximized
around 30 DIV \label{fig:peak-of-reverberation}}
\end{figure}

After plating, spontaneous activities are observed in about a week
in vitro. Such activities become synchronized into network bursts
around 10 DIV and show reverberations after 15 DIV. The number of
peaks per burst reaches a maximum around 30 DIV as shown in Fig.~\ref{fig:peak-of-reverberation}
and falls back to one without reverberation after 40 DIV. 
 It has been observed that the reverberatory bursts during the intermediate
DIV can be divided into two phases \citep{gandolfo_tracking_2010}:
a propagation phase where the channels are activated sequentially
and diffusively and a reverberation phase where the firings of the
neurons are seemingly random and more decoupled. Such division was
confirmed with CAT observation. As evident from streches of the CATs
shown in the insets of Fig.~\ref{fig:Predictability-Histogram}b
for a reverberatory burst, the propagation is indeed more prominent
for the initiating peak of spike rate (blue trajectory) and reduces
to a lingering (green) trajectory soon after. However, as the network
reverberates, the CAT gradually regains its propagating sweeps until
the end of the burst (magenta trajectory). 

\begin{figure}
\includegraphics[width=1\columnwidth]{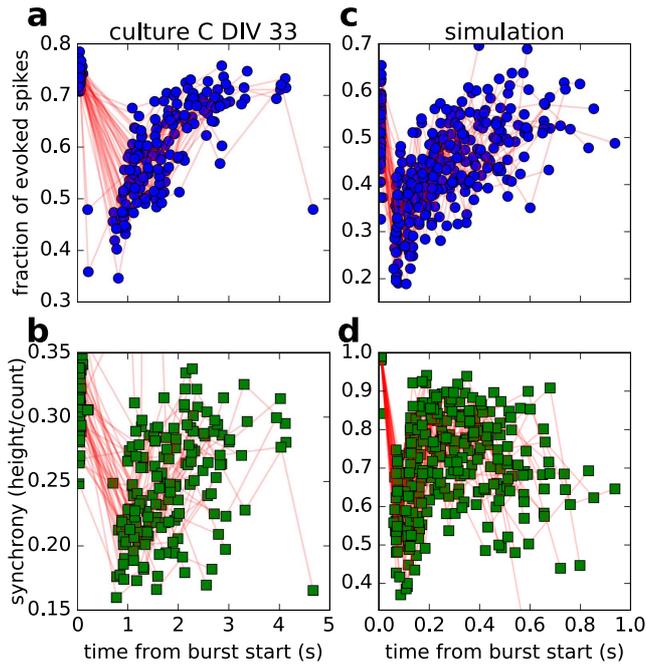}\caption{\textbf{a} Fraction of evoked spikes in detected activity peaks for
reverberatory bursts of 33 DIV recording of culture C against the
peak times relative to the start of the bursts. \textbf{b} Peak synchrony
defined as the height of a peak over its spike count. The faint lines
connect activity peaks in a burst in sequence. \textbf{c} and \textbf{d}
are corresponding results for fraction of evoked spikes and synchrony,
respectively, from simulations. \label{fig:Fraction-of-spread}}
\end{figure}

The factors driving the spiking activity of a neuron during a bursting
event include the synaptic action spreading from its presynaptic neurons
and the spontaneous activation driven by its own neuronal or synaptic
noises. To identify the dominating factor contributing to a spike,
we use the simple linear-spread diffusive model (\ref{eq:linear-spread-diffusive})
parametrized with a spreading speed, which can be determined by a
maximum likelihood method for each recording as documented in Table
\ref{tab:selected-cultures}. While a more sophisticated propagation
model might produce a better match to the observed behavior, the added
complexity is not expected to change our conclusions qualitatively.
Using the propagation model (\ref{eq:linear-spread-diffusive}), we
classify spikes into evoked spikes and spontaneous spikes. We then
determine their ratio for all rate peaks of a burst. The results of
evoked-spike fractions plotted in Fig.~\ref{fig:Fraction-of-spread}a
for 33 DIV recording of culture C show an increase in the fraction
of evoked spikes as the network reverberates. 
To characterize how synchronous the spikes within an activity peak
are, we normalize each rate peak $i$ with its spike count $n_{i}$
and use the normalized height $h_{i}/n_{i}$ to quantify the synchrony.
In Figure~\ref{fig:Fraction-of-spread}b, the synchrony of the activity
peaks is plotted against the time of the peaks relative to the start
of the bursts. While the synchrony data is more disperse, we can see
an upward trend following the time course of the bursts. This demonstrates
a correlation between the activity spreading and synchrony of the
spikes. The result may not be a surprise considering the activity
spreading through synaptic action following presynaptic spikes is
how neurons can communicate and should help to orchestrate the synchronous
activity.

\begin{figure}
\includegraphics[width=1\columnwidth]{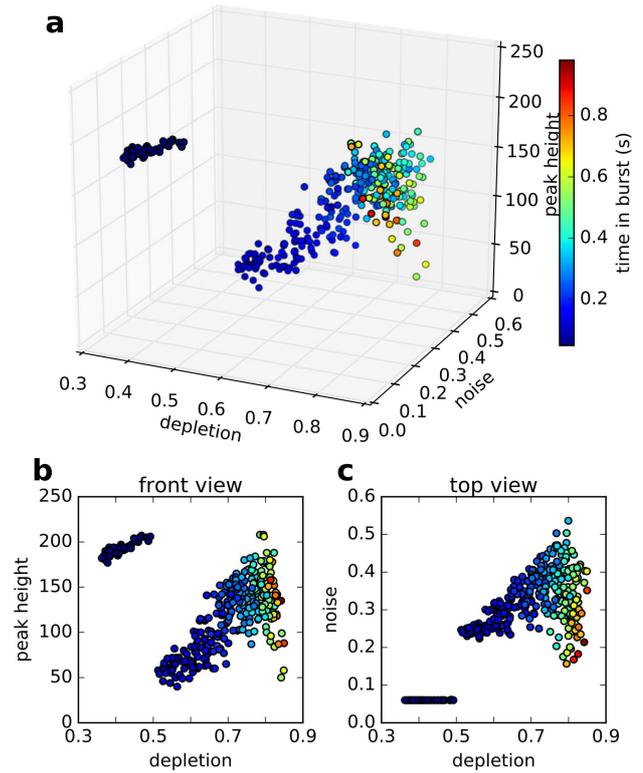}

\caption{\textbf{a} Scatter plot for correlation of height of activity peaks
with the system level of noise, represented by residual calcium concentration
({[}Ca$^{2+}${]}), and the depletion of synaptic resources, represented
by the super inactive state of neural transmitters. \textbf{b} Front
view of the 3D plot. \textbf{c} Top view of the 3D plot \label{fig:height-noise-depletion}}
\end{figure}

To further clarify the synaptic dynamics contributing to the increasing
dominance of the evoked spikes over spontaneous ones during a reverberatory
burst, we turn to our simulations that produce qualitatively similar,
reverberatory bursts with the increasing height of activity peaks
in the spike-rate, time histogram over the bursts as shown in Fig.~\ref{fig:Predictability-Histogram}c.
With a simulated system, the full set of dynamic variables are available
for analysis. We identify two factors of relevance in determining
the peak height or the synchrony of the reverberation from our simulations:
Firstly, the residual calcium concentration controls the rate of asynchronous
release at synapses in the model and represents the strength of an
internal \emph{noise} of the neurons. Secondly, the inactive and super-inactive
states featured in the model take up the neural transmitters as they
are activated and represent the \emph{depletion} of synaptic resources.
We correlate the system average of these two factors with the height
of activity peaks in a 3D scatter plot for all peaks of the simulated
recording as shown in Fig.~\ref{fig:height-noise-depletion}. 
 From the projection Fig.~\ref{fig:height-noise-depletion}b, we
see that depletion, which increases during a burst, correlates positively
with an increase of the peak height and thus the synchrony of the
spikes. On the other hand, the noise factor represented by residual
calcium, as shown in Fig.~\ref{fig:height-noise-depletion}c, is
initially pumped up by the spiking activity of a burst, reaching a
maximum about half way through the burst, and decreases afterward
due to the lengthening intervals between the reverberation peaks until
the end of the burst.

\begin{figure}
\includegraphics[width=1\columnwidth]{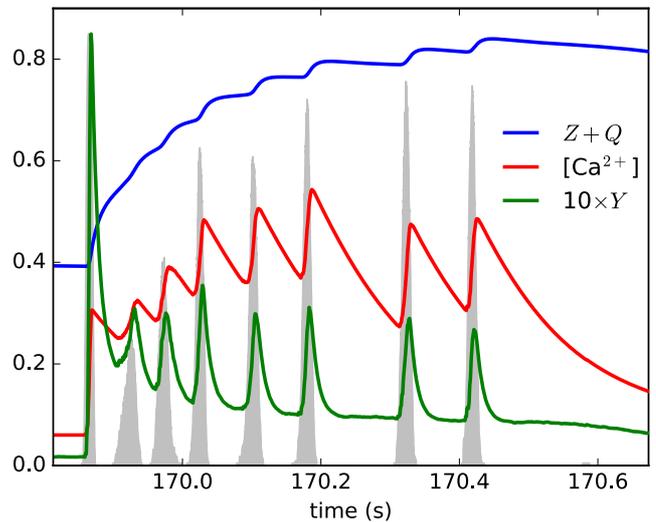}

\caption{Average levels of residual calcium concentration (middle curve), depleted
neural transmitter (upper curve), and active neural transmitter fractions
(lower curve, 10-time magnified) of the system over the course of
the simulated burst of Fig. \ref{fig:Predictability-Histogram}c.
The shaded area shows the time-histogram of the burst \label{fig:mean-variables-over-burst}}
\end{figure}

The detailed dynamics of different factors can be further analyzed
in a simulation. In Fig. \ref{fig:mean-variables-over-burst}, we
plot the residual calcium concentration, depleted neural transmitter
fraction ($Z+Q$), and the active neural transmitter fraction over
the very burst shown in Fig. \ref{fig:Predictability-Histogram}c.
 Taken from the computational model, the depletion of neural transmitters
to the $Z$ and $Q$ states is driven by the activated transmitters
$Y$ from the spiking activity. The spiking activity also increases
the level of residual calcium, which controls the noisy, asynchronous
releases of the neural transmitters leading to the reverberation in
a burst. Apart from making the role of synchronous releases more important
in the activation of neurons, the depletion also leads to longer recovery
time of neural transmitters. Coupling with the rapid decay of calcium
between the reverberation peaks, this leads to the lowing in the mean
level of residual calcium towards the end of a burst.

\section{Discussion}

In the current study, we use a high-density MEA system to investigate
the physical mechanisms underlying the morphological richness of reverberatory
neural bursts. Our simple linear-spread diffusive model allows a classification
of the spikes as well as an inference of the propagation speed of
synaptic activities. The change of the predictability of spikes allows
us to detect the change in the propagation behavior during a burst
as shown in Fig.~\ref{fig:Fraction-of-spread}. However, the traveling-wave-like
sweep of activity, especially for the initiation of a burst, is not
diffusive. A more sophisticated model will be required if one would
like to have a more faithful capture of such dynamics. Nonetheless,
the method of inference for the model parameters using individual
spikes as demonstrated remains applicable. The method is enabled by
our use of high-density MEA and does not resort to data reduction
before inferring the propagation dynamics. That is, each spike has
a direct contribution to the resolution of the spreading speed and
the method can potentially be used to resolve more complex dynamics
of the system.

The finding from our analysis of the simulated system suggests an
interesting phenomena, which we call \emph{depletion-enhanced synchronization},
at play in the cultured network with the reverberatory bursts. In
such a burst, the initiation activity is a fast sweeping wave of propagating
spikes across the network that is well synchronized. This activity
produces a significant amount of residual calcium, promoting noisy
asynchronous releases, and prompting the spontaneous firing of the
neurons that results in the subsequent reverberation of the burst.
Initially, the spontaneous spikes are more or less independent and
the heterogeneity in the neurons and their connectivity makes the
spike-rate peaks broad and less synchronous. However, as the neural
and synaptic resources are increasingly depleted by the continuing
spiking activity of the burst, it becomes harder for the neurons to
fire independently and they thus increasingly rely on the synchronous
releases triggered by the firing of their presynaptic neurons to help
them cross the firing threshold. Such mechanism accounts for the observed
increase of evoked spikes and the synchrony in Fig.~\ref{fig:Fraction-of-spread}
and may be a general mode of operation for other complex systems.

The synchronized network activities observed in our cultures seem
to be similar to the switching between Up and Down states as observed
in other neuronal network preparations \citep{maclean_internal_2005,holcman_emergence_2006,johnson_development_2007}.
However, since our measurements are carried out on MEA, records of
the membrane potentials are not available to verify these states.
It is known that activities similar to what we reported here can also
be induced in acute slice \citep{czarnecki_network_2012} when inhibitory
interactions are blocked. Presumably, there are too many recurrent
connections in our cultures which might correspond to the pathological
condition during epilepsy \citep{mccormick_cellular_2001}.

In the computational model, the active state is initially stabilized
by the residual calcium which promotes the asynchronous releases intrinsic
to the neurons, and later revitalized by the synaptic couplings of
the network. The role of calcium in the reverberation was implicated
by \citet{lau_synaptic_2005} and we chose to implement the model
by \citet{volman_calcium_2007} based the similarity between the firing-rate
time histograms it produces and those were seen in our experiments.
Alternatively, NMDA receptors have been proposed to play a role in
persisting a burst \citep{wang_synaptic_1999,wang_synaptic_2001}.
It will be interesting to see in future studies what difference in
the bursting morphology will result from an NMDA receptor based model.

Finally, we note that while synchrony is often associated with coherence,
it actually reduces the diversity in the possible dynamics of a system.
In the reverberatory bursts that we focused on, the synchrony results
from the depletion of synaptic resources and precedes the termination
of the burst. This parallels the recent findings in epilepsy that
increasing synchrony can be observed towards the end of seizures \citep{lehnertz_synchronization_2009,jiruska_synchronization_2013}.
Our results may suggest a possible mechanism for such phenomena for
systems of similar episodic dynamics.
\begin{acknowledgements}
This work has been supported by the Ministry of Science and Technology,
Taiwan, Republic of China under grand no. 102-2112-M-001 -009 -MY3,
and Physics Division, National Center for Theoretical Sciences, Taiwan,
Republic of China.
\end{acknowledgements}

\end{document}